\documentclass[
    ,final            
  ]
  {aipproc}

\layoutstyle{6x9}

\begin{document}

\title{The Rise and Fall of Pentaquarks in Experiments.}


\author{Reinhard A. Schumacher}{
 address={Department of Physics, Carnegie Mellon University,
 Pittsburgh, PA 15213}}

\begin{abstract}

Experimental evidence for and against the existence of pentaquarks has
accumulated rapidly in the last three years.  If they exist, they
would be dramatic examples of hadronic states beyond our well-tested
and successful particle models.  The positive evidence suggests
existence of baryonic objects with widths of at most a few MeV, some 
displaying exotic quantum numbers, such as baryons with strangeness $S
= +1$.  The non-observations of these states have often come
from reaction channels very different from the positive evidence
channels, making comparisons difficult.  The situation has now been
largely clarified, however, by high-statistics repetitions of the
positive sightings, with the result that none of the positive
sightings have been convincingly reproduced.  The most recent
unconfirmed positive sightings suffer again from low statistics and
large backgrounds.  It seems that a kind of ``bandwagon'' effect led
to the overly-optimistic interpretation of numerous experiments in the
earlier reports of exotic pentaquarks.

\end{abstract}

\maketitle


\section{INTRODUCTION} 

This talk will discuss the experimental evidence for and against
pentaquarks, in particular the ``exotic'' baryonic states that have
charge and flavor quantum numbers that require a minimal valence quark
configuration of four quarks and one anti-quark.  The recent work in
this field began with the first report~\cite{Nakano:2003p}, at the
2002 PANIC conference, of a narrow $S = +1$ at 1540 MeV decaying to
$K^+ n$.  There are older chapters in the search for positive
strangeness ``$Z$'' resonances~\cite{oldpdg} and of charmed strange
pentaquarks~\cite{Ashery:1997} that we will omit here.  In the three
years since the announcement of a positive strangeness baryon, named
the ``$\Theta^+$'', an enormous amount of work has been done, seemingly by
every particle physics collaboration in the world, to seek evidence
for narrow pentaquark states.  This led to the sightings for states
identified as the $\Theta^+$, the $\Xi_5^{--}$, the $\Theta_c^0$, and
the $\Theta^{++}$.  After an initial flurry of positive reports for
these states, null results, that is, searches that led to no observed
narrow exotic structures, started to dominate the field.  As I will
try to show below, from the perspective of the 2005 PANIC conference,
there was a certain sociological ``bandwagon effect'' in 2003 and the
first half of 2004 in which numerous groups rushed be part of the wave
of positive sightings.  The bandwagon lost momentum when measurements
repeated with higher statistics gave negative results, and when
numerous negative searches emerged in previously unexplored channels.
While over 50 experimental papers discussing pentaquark searches can
now be found in the literature, over the same time period there have
been over 550 theoretical papers.  Thus, the impact of pentaquark
searches has been to renew interest in models of QCD which try to
address the fundamental question of why nature prefers only the
conventional 3-quark baryonic states and not more complex
configurations such as pentaquarks.

What are pentaquarks?  QCD does not explicitly forbid baryons with
four quarks and an antiquark, or mesonic-like states two quarks and
two anti-quarks.  This was discussed in the context of bag models, for
example, by Jaffe~\cite{Jaffe:1976ig} and deSwart {\it et
al.}~\cite{deSwart:1980db}.  In soliton models there were early
discussions by, among others, Kopeliovich~\cite{kopeliovich},
Chemtob~\cite{chemtob}, and Walliser~\cite{walliser}.  However, the
recent interest in pentaquarks stems from bold predictions made by
Diakonov, Petrov and Polyakov~\cite{dpp} in the context of a
chiral-quark soliton model for an anti-decuplet of pentaquark states.
They predicted a narrow $(\Gamma_{\Theta^+} \sim 15$ MeV) $uudd\overline{s}$
state, the $\Theta^+$, close to $M = 1530$ MeV.  The same model
predicted seven non-exotic pentaquark baryons which could behave like
ordinary $N^*$ states or hyperons.  The model also predicted a set of
cascade-like ($S=-2$) states near 2070 MeV, two of which, the
$\Xi^{--}_5$ and the $\Xi^+_5$ had the charge-flavor--exotic
structures $ddss\overline{u}$ and $uuss\overline{d}$, respectively. In
this model, the pentaquarks emerge as rotational excitations of the
soliton, with $J^P=\frac{1}{2}^+$.

The initial experimental reports of states corresponding, perhaps, to
this specific anti-decuplet prediction led to extensive theoretical
re-evaluation of QCD models to shed additional light on pentaquark
physics.  For example, in a quark-model approach Jaffe and
Wilczek~\cite{Jaffe:2003sg} explored the possibility of di-quark
attraction strong enough to cause new stable hadronic structures.  In
that model, the pentaquarks consist of two bosonic $(ud)$ diquark
pairs coupled via $L=1$ to an anti-quark, with the lowest states
having $J^P=\frac{1}{2}^+$.  An octet of di-quark pentaquarks was predicted to
accompany the anti-decuplet, and it was suggested that charm and
bottom analogs to the $\Theta^+$ might also be stable against strong
decay.

The lattice community tried to find evidence for pentaquark
structures, and results from a total of 10 groups have
appeared~\cite{lattice}.  In all those studies, the crucial point was
to distinguish a simple $KN$ continuum scattering state from a
``bound'' $\Theta^+$ pentaquark structure.  The results were
conflicting, with equal numbers of results reporting a pentaquark
structure as not, and with disagreement among the affirmative results
regarding the spin and parity of the ground state.

\section{EXPERIMENTAL EVIDENCE}

\begin{figure*}
\centering
\resizebox{0.85\textwidth}{!}{\includegraphics[angle=90]{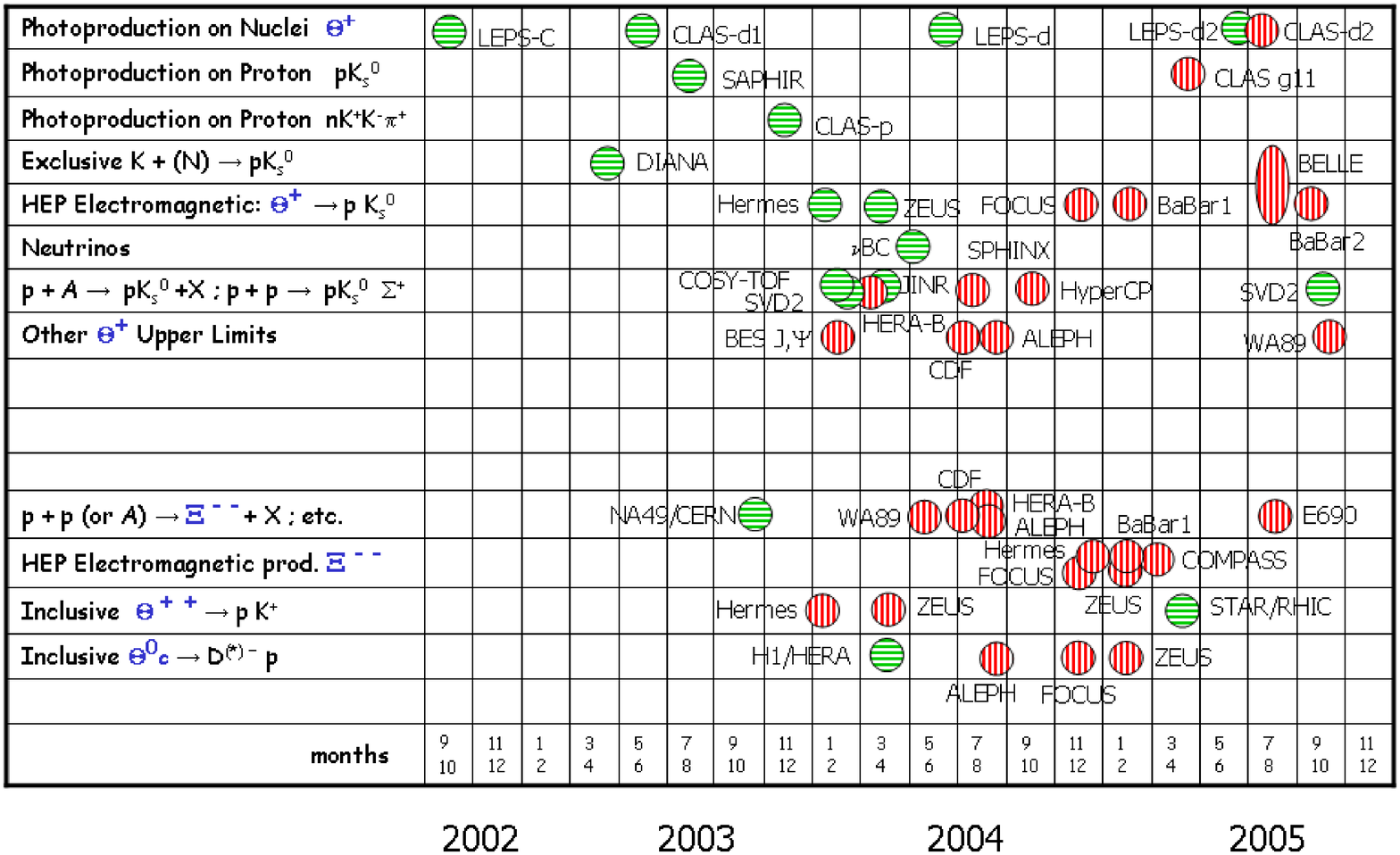}}
\caption{
Time line of the experimental pentaquark searches from 2002 to 2005.
Horizontal bars (green circles) designate claims of sighting, while
vertical bars (red circles) designate reported non-observations. 
}
\label{fig:timeline} 
\end{figure*}

A way of summarizing the various categories of pentaquark searches
since October 2002 is given in Fig.~\ref{fig:timeline}.  The table
groups experiments by reaction type, and uses green (horizontal
bars) or red (vertical bars) circles to show the experiments in each
category that found positive or negative evidence, respectively, for
pentaquarks.  For example, in the first row are the experiments done
with few-GeV real photons on deuterium or carbon.  Four positive
$\Theta^+$ pentaquark sightings are entered, but then a recent
negative result completes the row.  The negative result in this
reaction category was a high-statistics repetition (from CLAS at
Jefferson Lab (CLAS-d2)~\cite{baltzell}) of one of the previous
positive sightings (also from CLAS at Jefferson Lab
(CLAS-d1)~\cite{Stepanyan:2003qr}).  The structure of the table is
meant to suggest that pentaquarks in this category of reaction have
likely been ruled out.  As another example, the one positive
$\Xi^{--}_5$ measurement from NA-49/CERN~\cite{Alt:2003vb} is
followed, on two rows of the table, by at least 10 negative results
from either hadronic or electromagnetically induced reactions.

The intent of the table in Fig.~\ref{fig:timeline} is also to show how
a kind of ``bandwagon'' effect took place starting in late 2002, with
a succession of positive pentaquark sightings from diverse
collaborations.  By the middle of 2004 a total of 14 positive
sightings were reported.  There was concern about the consistency of
these reports, especially the somewhat discrepant reported mass
values, and the generally poorly-understood backgrounds in the
reported mass spectra.  The rush of positive sighting then stopped,
and from the middle of 2004 until late 2005 most of the reports were
negative. As seen in the table, in many cases one can say that a given
negative result essentially repudiates a specific previous positive
sighting.  Few experiments have been repeated under exactly identical
experimental conditions, so there is room for continued debate in some
cases.  The remaining unchallenged positive findings (rows not ending
in a red (vertical bars) marker) are becoming harder to reconcile.

This table is not exhaustively complete.  For example, more
experiments than are shown reported \emph{not} seeing a $\Theta^{++}$
state, though rigorous upper limits were seldom given.  Also, we omit
the phenomenological studies that reanalyzed existing $KN$ phase shift and
scattering data to infer a maximum allowable width for the $\Theta^+$.
These led to estimates of $\Gamma_{\Theta^+} \simeq 1$ MeV for the 
allowable total width of the
$\Theta^+$~\cite{Arndt:2003xz}~\cite{Gibbs:2004ji}~\cite{Cahn:2003wq}.

In the category of real GeV photons on  nuclei,
the initial report from LEPS/SPring-8~\cite{Nakano:2003p} used a carbon
(scintillator) target; a subtle Fermi-motion correction was needed
since only a $K^+$ and $K^-$ were detected in the reaction $\gamma (N)
\rightarrow K^+ K^- X$.  A follow-up experiment from LEPS/SPring-8 on 
deuterium~\cite{nakano-d} was performed under the same conditions as
the carbon experiment.  More evidence for the $\Theta^+$ formation was
shown at conferences, but that result is unpublished.  The LEPS
collaboration also reported a weak positive signal, from the same data
set, in the exclusive channel $\gamma d \rightarrow \Lambda(1520) ~
\Theta^+$~\cite{nakano-d2}.  The identified $\Lambda(1520)$ tagged the
strangeness of the $\Theta^+$, but the signal sat on a large and
steeply-sloped background.  Stronger positive evidence for a
$\Theta^+$ was reported by CLAS/JLab~\cite{Stepanyan:2003qr} in an
exclusive measurement on deuterium, $\gamma d \rightarrow p K^+ K^-
(n)$.  A 4.6 to 5.8 $\sigma$ signal was seen above a large background
that was difficult to estimate quantitatively due to nuclear final
state interaction effects.  Both LEPS and CLAS suffered from unknown
or poorly-known background distributions under their resolution- and
statistics-limited signal peaks.

CLAS at Jefferson Lab repeated the measurement on deuterium in early
2004, and reported results in mid-2005~\cite{baltzell}.  With six
times the statistics, no $\Theta^+$ peak was seen in the $\gamma d
\rightarrow p K^+ K^- (n)$ reaction, and a model-dependent upper limit
of 4 to 5 nanobarns was set for a state at 1.54 GeV.  The previous
CLAS result, when fitted with a luminosity-scaled background shape
from the higher statistics run, was reduced in significance to about
$3\sigma$.  CLAS also did not see a signal in the $\Lambda(1520)\Theta^+$
final state, albeit the kinematic conditions were not the same as
those at LEPS.  On balance, there is no good strong experimental
evidence left to suppose the existence of a $\Theta^+$ produced in
these channels.

In exclusive production by GeV-scale real photons on the
proton, excitement was generated by the SAPHIR Collaboration at
Bonn~\cite{Barth:2003es} when they reported a 300 nanobarn positive
signal in the reaction $\gamma p \rightarrow \overline{K^0} ~ \Theta^+ 
\rightarrow \pi^+ \pi^- K^+ (n)$.  The strangeness of the decaying 
$K^0_S$ was indeterminate, but the narrowness of the signal, $\Gamma < 25$ MeV,
was believed to signal formation of the pentaquark. This same reaction
was tested with much higher statistics at the same kinematics by the
CLAS Collaboration~\cite{Battaglieri:2005er}.  No $\Theta^+$ signal
was found, and an integrated upper limit of 0.8 nb was set at 1.54
GeV, with similar results over a wide range of mass values and
production angles.

A second channel of exclusive production on the proton was reported by
CLAS~\cite{Kubarovsky:2003fi}, in $\gamma p \rightarrow \pi^+ K^-
K^+ (n)$.  The neutron was detected by missing mass, the pion was
constrained to go into the forward hemisphere in the center of mass,
and the $K^+$ from the putative pentaquark decay was constrained to be
in the backward hemisphere.  This combination of cuts was thought to
enhance diffractive production of a high mass non-strange nucleon
resonance which could then decay to a $K^-$ and a $\Theta^+$.  A
$7.8\sigma$ signal was reported.  So far, this measurement has not
been repeated at any other lab.  It is perhaps the most convincing
remaining candidate as of the beginning of 2006.

The ideal way to make a $\Theta^+$ pentaquark that has a mass near
1.54 GeV is to scatter $K^+$ particles of about 430 MeV/c momentum
from neutrons.  Review of the low energy kaon scattering phase shifts
revealed no signatures, but left open the possibility of excitation of
a small, less than 1 MeV wide
state~\cite{Arndt:2003xz}\cite{Gibbs:2004ji}\cite{Cahn:2003wq}.  The
DIANA collaboration at ITEP used a slowing 850 MeV/c $K^+$ beam in a
xenon bubble chamber to report a narrow signal at 1.54
GeV~\cite{Barmin:2003vv}.  The detected final state of $K^0_S p$ had
no well-defined strangeness, but the initial $K^+$ did define the
strangeness.  In recent years there have been no high intensity low
momentum kaon beams in the world to repeat this kind of measurement.
To evade this problem, the Belle Collaboration at KEKB, searched for
hadronic events stemming from identified $K^+$'s, produced in $e^+e^-$
collisions, that interacted in detector elements near the collision
point~\cite{Abe:2005gy}.  Their sensitivity was slightly better than
that of the DIANA experiment, but they did not confirm the previous
measurement.  An upper limit given in terms of the width of the state
was reported, $\Gamma_{\Theta^+} < 0.64$ MeV.

A second nearly ideal way to make a $\Theta^+$ pentaquark is in the
reaction $p p \rightarrow \Sigma^+ ~ \Theta^+$, where the pentaquark
decays to $K^0_S p$.  This avenue was followed by the TOF
Collaboration at COSY/J\"{u}lich which
reported~\cite{Abdel-Bary:2004ts} a 4 to 6 $\sigma$ signal in this
channel using a non-magnetic time of flight spectrometer.
Unfortunately, their experimental background under the signal, like in
many of the other positive sightings, was not calculable, and so had
to be fitted with a polynomial.  This leads to concern about the
reliability of the signal-to-background estimation.  Also, there was
no hint of the $\Theta^+$ band in a $\Sigma^+ p K^0$ Dalitz plot
analysis.  This measurement has not been repeated yet at COSY or any
other laboratory, so it stands, at the present time, as a surviving
candidate.

At much higher energies, one can consider pentaquark production in the
fragmentation of quark systems from various targets.  The HERMES
Collaboration at HERA reported ~\cite{Airapetian:2003ri} a
$\Theta^+$ signal from quasi-real photons ($Q^2 \sim 0$) produced in
positron scattering at 27 GeV on deuterium.  It was a statistically
weak signal, relying on background estimation from a standard Monte
Carlo fragmentation model, but also invoking the presence of a number
of poorly-characterized excited hyperon resonances, in addition to the
putative $\Theta^+$.  At the other end of the $Q^2$ continuum, a
$\Theta^+$ signal was reported by the ZEUS
Collaboration~\cite{Chekanov:2004kn} at HERA from $ep$ collisions near
a c.m. energy of 310 GeV. A narrow bump appeared at 1.52 GeV above a
quark fragmentation Monte Carlo background estimation, but only for
$Q^2 > 20$ (GeV/c)$^2$.  The signal was visible in both the $K^0_S p$
and the $K^0_S \overline{p}$ combinations, which, while neither tags
the strangeness of the signal uniquely, does hint at the formation of
both $\Theta^+$ and $\overline{\Theta^+}$.  Unfortunately, the signal
resides very near the steeply-sloped phase-space background for the
detected final state, such that the significance of the signal hinges
crucially on the reliability of the background Monte Carlo model.
Furthermore, a second bump near the putative $\Theta^+$ was seen, and
this was ascribed to a poorly-known $\Sigma$ hyperon resonance, and such
hyperon bumps are otherwise unseen in high energy experiments.

These two experiments in high energy lepton scattering can be
contrasted with the much higher statistics measurements from the BaBar
Collaboration at SLAC~\cite{Gotzen:2005vm}.  In that measurement, beam
halo electrons and positrons scattered from the beam pipe surrounding
the interaction region, resulting in nuclear scattering of the leptons
from beryllium.  Such events are dominated by the lowest $Q^2$'s, and
thus are comparable to the kinematics of the HERMES measurement.
There was no hint of a $\Theta^+$ in the BaBar $K^0_S p$ spectrum,
contradicting the previous experiment.  The comparison of BaBar to
ZEUS is less significant since the latter result was for $Q^2 >
20$ (GeV/c)$^2$; indeed, ZEUS saw no signal at lower $Q^2$.  We also
note that BaBar did not see a signal for the $\Theta^+$ or any of the
other pentaquark states in $e^+e^-$ collisions not related to ``beam
pipe'' scattering, and set upper limits about an order or magnitude
below the production of ordinary baryons near the candidate mass
values~\cite{Berger-Hryn'ova:2005vq}.

Mining old data for new phenomena, five neutrino bubble chamber
experiments from CERN and Fermilab were combined to report a narrow
peak of a few dozen $K^0_S p$ events at 1.53
GeV~\cite{Asratyan:2003cb}.  There was a clear excess of events above
the mixed-event background, not only at the location of the putative
pentaquark but also at higher masses.  Thus, it is possible that the
estimated background was not understood well enough to reliably claim
the presence of a specific new signal at 1.53 GeV.

In scattering hadronic probes at high energy from nuclear target, a
positive signal for the $\Theta^+$ decaying to $K^0_S p$ was reported
by the SVD Collaboration, using 70 GeV protons in a fixed-target
arrangement at a c.m. energy of about 11.5 GeV.  Their initial
report~\cite{Aleev:2004sa} was supported by a more
recent~\cite{Aleev:2005yq} detailed analysis which increased their
pentaquark signal by a factor of about 8.  The signal in the $K^0_S p$
mass spectrum comprises about 300 events per channel, with a
substantial background under the peak that was modeled using mixed
events and standard Monte Carlo.  The statistics of this
measurement must be contrasted to the results of WA-89 Collaboration
for the scattering of a 340 GeV/c $\Sigma^-$ beam from carbon and
copper~\cite{Adamovich:2005ns}.  At these high energies it is hard to
imagine that the beam energy, probe, or target material could make
much difference to the production of pentaquarks.  Thus, the
featureless 40,000 counts per channel in the WA-89 spectrum strongly
suggests that hadronic production of the $\Theta^+$ has not occurred.
The SPHINX Collaboration at IHEP had also looked using 70 GeV protons
on carbon for a $\Theta^+$ signal with a null
result~\cite{Antipov:2004jz}. They were able to reconstruct decay
channel $K^0_S p$, but also had results for $K^0_L p$ and $K^+n$.
Their upper limits were in the range of 30 nb/nucleon, and a
production ratio of $\Theta^+ ~
\overline{K^0}$ to $\Lambda(1520) K^+$ of less than 0.02.  A further
negative result for 800 GeV protons on a carbon target was reported by
HyperCP at Fermilab~\cite{Longo:2004gd}.  Their upper limit was given
as less than $0.3\%$ $\Theta^+$ production of all reconstructed $K^0_S
p$ events.  Finally, a negative result was reported by
HERA-B~\cite{Abt:2004tz} for the interaction at 41.6 GeV c.m. energy
of protons on several nuclear targets.  They set modest upper limits
for $\Theta^+$ production of less than 16 $\mu$b/N and less than about
$12\%$ relative to the $\Lambda(1520)$.  Thus, despite the recent
positive result reported by SVD-2, the evidence is greatly against the
production of $\Theta^+$ pentaquarks in high energy hadronic
production.

Further non-observations of the $\Theta^+$ were reported for $J/\psi$
decays involving $\Theta^+ \rightarrow K^0_S p$ from
BES~\cite{Bai:2004gk}, for $p\overline{p}$ collisions from
CDF~\cite{Litvintsev:2004yw}, and from events in $e^+e^-$ collisions
at the $Z$ pole from ALEPH~\cite{Schael:2004nm}.  The significance of
these results in relation to the positive observations at low energies
is difficult to estimate, since the production mechanism of exotic
pentaquarks is, well, exotic.  Nevertheless, these results add some
weight to the conclusion that these states have in fact not been seen.

In the theoretical models mentioned earlier~\cite{dpp}
\cite{Jaffe:2003sg},
the other exotic pentaquarks that were predicted were two cascade-like
($S=-2$) states in the mass range of 1.75 to 2.07 GeV.  The NA-49
Collaboration at CERN found evidence~\cite{Alt:2003vb} for narrow
states at 1.86 GeV in $pp$ collisions at 17 GeV c.m. energy, detected
via decays such as $\Xi^{--}_5 \rightarrow \Xi^- \pi^-$.  (Note that
the Particle Data Group calls this state the $\Phi(1860)$.)  The close
agreement of the predictions with this experimental result was very
exciting, especially in view of the previous seemingly accurate
predictions of the mass of the $\Theta^+$.  However, a deluge of
contrary experimental results followed this one positive claim for a
cascade pentaquark.  At least ten results have been released which
repudiate the existence of the states seen by NA-49: from
WA-89~\cite{Adamovich:2004yk}, CDF~\cite{Litvintsev:2004yw},
HERA-B~\cite{Abt:2004tz}, ALEPH~\cite{Schael:2004nm},
HERMES~\cite{Airapetian:2004mi}, FOCUS~\cite{Stenson:2004yz},
BaBar~\cite{Aubert:2005qi}, ZEUS~\cite{Chekanov:2005at},
COMPASS~\cite{Ageev:2005ud}, and
E690/Fermilab~\cite{Christian:2005kh}.  In general, the negative
reports have much higher statistics than the original positive report,
and came from $pp$, $ep$, $pA$, and $\gamma A$ experiments.  Upper
limits relative to production of the well-known $\Xi(1530)$ state were
typically an order of magnitude below the NA-49 observation.  Thus,
there is really no chance left that these exotic pentaquarks
candidates exist at measurable production levels.

The H1 Collaboration at HERA reported evidence~\cite{Aktas:2004qf} in
$ep$ collisions for a narrow anti-charmed baryon dubbed the
$\Theta_c^0$, which would have a minimal quark configuration of
$uudd\overline{c}$.  The state had a mass of 3.099 GeV, was as narrow
as the experimental resolution, and detected via the decays $D^{*-} p$
and $D^{*+} \overline{p}$.  It was reported to contribute about $1\%$
of the $D^*$ production rate in DIS.  As in many of the other
pentaquark searches, the signal comprised two or three ``high''
channels above a very substantial background.  A 5.4$\sigma$
significance was claimed for the high-count region.  Much higher
statistics negative results were subsequently reported in the same
final states by ZEUS~\cite{Gladilin:2004qg}, FOCUS~\cite{Link:2005ti},
and ALEPH~\cite{Schael:2004nm}.  No other positive sightings have come
out.

If the $\Theta$ pentaquark were an isovector object, not the isoscalar
predicted in most models, then other charge states exist, such as a
$\Theta^{++}$.  The decay of this state to $K^+ p$ is especially easy
to look for.  Whereas many non-sightings have been mentioned in the
pentaquark literature, a recent result from the STAR Collaboration at
RHIC claims~\cite{Huang:2005nk} a narrow state decaying to $K^+ p$ in
deuteron-gold collisions at 200 GeV $NN$ c.m. energy, with a
4.2$\sigma$ significance.  The state sits atop a very large but smooth
mix-event background, but unfortunately also sits next to an equally
large bump that is attributed to $K/\pi$ particle identification
errors.  Confirmation of this structure and its interpretation are
clearly needed.

\section{CONCLUSIONS}

In summary, after three years of intense activity, pentaquarks have
come and gone.  In this paper/talk we have not had time or space to
dwell on many specifics of any of the experiments, but the overall
trends are very consistent.  Most of the positive sightings have been
contradicted or placed in doubt by better measurements.  The remaining
candidates have no common thread to unify their phenomenology.  Most
important, no single truly convincing positive measurement claim has
appeared.  Recent new candidates suffer (again) from low statistics
and poorly-understood backgrounds.  As was shown in this paper, there
was a rush of positive sightings between late 2002 and mid-2004.  From
that time forward almost all experiments showed null results for
detection of $\Theta$, $\Xi_5$ and $\Theta_c$ pentaquarks.  Thus, one
can conclude that a ``bandwagon'' rush of over-optimistic positive
sightings was in effect initially, but now the lack of convincing
evidence for narrow exotic pentaquarks is overwhelming.

{\bf Acknowledgments:}
The collection of information for this talk and paper was facilitated by 
discussions with V. Burkert, C. Meyer, D. Tedeschi, K. Hicks, and P. Stoler.



\end{document}